\documentclass[final,5p,times,twocolumn, numcompress]{elsarticle}

\usepackage{amssymb}
\usepackage{lipsum}
\usepackage{amsmath}
\usepackage{amsfonts}
\usepackage{graphicx}
\usepackage{array}
\usepackage[table,xcdraw]{xcolor}
\usepackage{colortbl}
\usepackage{float}
\usepackage{natbib}
\usepackage{algpseudocode}
\usepackage{multirow}
\usepackage{booktabs}
\usepackage{adjustbox}
\usepackage{svg}
\usepackage{caption}
\usepackage{subcaption}
\usepackage{amsmath}
\usepackage[ruled]{algorithm2e}
\usepackage{nicematrix}
\usepackage{pdflscape}
\usepackage{orcidlink}

\definecolor{lightgray}{RGB}{240, 240, 240}
\definecolor{steelblue}{RGB}{52,138,189}
\definecolor{cite-blue}{HTML}{008ADA}
\definecolor{gainsboro229}{RGB}{229,229,229}
\definecolor{tablecolor}{RGB}{49, 131, 245}

\definecolor{tab0}{HTML}{B0DBF1}
\definecolor{tab1}{HTML}{EFF8FC}
\definecolor{tab2}{HTML}{DEF0F9}
\definecolor{tab3}{HTML}{CFE9F7}
\definecolor{tab4}{HTML}{BFE2F4}
\definecolor{tab5}{HTML}{EFF8FC}
\definecolor{tab6}{HTML}{DEF0F9}
\definecolor{tab7}{HTML}{CFE9F7}

\title{Pioneering Precision in Lumbar Spine MRI Segmentation with Advanced Deep Learning and Data Enhancement}

\author{Istiak Ahmed\fnref{1}\orcidlink{0009-0004-1374-6300}} 
 \ead{istiak.ahmed1@northsouth.edu}
\author{Md. Tanzim Hossain\fnref{2}\orcidlink{0000-0002-4776-7220}}
 \ead{tanzim.hossain@fau.de}
 
\author{Md. Zahirul Islam Nahid\fnref{1}\orcidlink{0009-0008-1523-1297}}

\author{Kazi Shahriar Sanjid\fnref{1}\orcidlink{0009-0001-6845-8881}}

\author{Md. Shakib Shahariar Junayed\fnref{1}\orcidlink{0009-0009-8755-6370}}

\author{M. Monir Uddin\corref{cor1}\fnref{3}\orcidlink{0000-0002-9817-6156}}
 \ead{monir.uddin@northsouth.edu}
  \cortext[cor1]{Corresponding author}
  
\author{Mohammad Monirujjaman Khan\fnref{1}\orcidlink{0000-0003-0779-8820}}
 \ead{}

\address[1]{Department of Electrical \& Computer Engineering, North South University, Dhaka-1229, Bangladesh}
\address[2]{Department of Computer Science, Friedrich-Alexander University, Erlangen, Nürnberg, 91054, Germany}
\address[3]{Department of Mathematics and Physics, North South University, Dhaka-1229, Bangladesh}
\makeatletter
\renewcommand*{\ps@pprintTitle}{}
\makeatother
\begin{document}
\begin{frontmatter}

\begin{abstract}

This study presents an advanced approach to lumbar spine segmentation using deep learning techniques, focusing on addressing key challenges such as class imbalance and data preprocessing. Magnetic resonance imaging (MRI) scans of patients with low back pain are meticulously preprocessed to accurately represent three critical classes: vertebrae, spinal canal, and intervertebral discs (IVDs). By rectifying class inconsistencies in the data preprocessing stage, the fidelity of the training data is ensured. The modified U-Net model incorporates innovative architectural enhancements, including an upsample block with leaky Rectified Linear Units (ReLU) and Glorot uniform initializer, to mitigate common issues such as the dying ReLU problem and improve stability during training. Introducing a custom combined loss function effectively tackles class imbalance, significantly improving segmentation accuracy. Evaluation using a comprehensive suite of metrics showcases the superior performance of this approach, outperforming existing methods and advancing the current techniques in lumbar spine segmentation. These findings hold significant advancements for enhanced lumbar spine MRI and segmentation diagnostic accuracy.

\end{abstract}


\end{frontmatter}

\section{Introduction}
\label{introduction}

Accurate segmentation of lumbar spine structures in medical images is crucial for diagnosing and treating various spinal conditions, particularly those associated with low back pain \cite{roudsari2010lumbar, sheehan2010magnetic}. Low back pain is a prevalent medical issue that affects a significant portion of the population, leading to disability and reduced quality of life \cite{katz2006lumbar}. With the advent of advanced imaging technologies such as Magnetic resonance imaging (MRI), it is now possible to acquire high-resolution 3D images of the spine \cite{MRI_richardson2005pharmaceutical}. MRI is particularly valuable in this context because it provides detailed images of soft tissues, bones, and nerves \cite{van2002mri}, making it an essential tool for identifying pathological changes in the lumbar spine that contribute to low back pain \cite{MRI_Detection_kirkham2006good}.

Manual segmentation of these images is labor-intensive and time-consuming \cite{patil2013medical}, necessitating the development of automated segmentation methods. Lumbar spine segmentation is important in diagnosing because it precisely delineates anatomical structures, such as vertebrae, the spinal canal, and intervertebral discs (IVDs) \cite{roudsari2010lumbar}. Accurate segmentation facilitates the identification of abnormalities \cite{spi1_yang2023rau,identifying_abnormalities_1_yanagisawa2023convolutional,identifying_abnormalities_2_narayan2023comprehensive}, assists in planning surgical interventions \cite{surgical_1_warfield1998real,Surgical_2_li2021medical}, and aids in monitoring disease progression and treatment response \cite{aggarwal2011role}.

Deep learning, particularly Convolutional Neural Networks (CNNs), has shown great promise in this domain, offering substantial improvements in segmentation accuracy and efficiency \cite{zhou2021review}. Automating the segmentation process using deep learning reduces the burden on radiologists \cite{cheng2021deep,radio_1_wataya2023radiologists, radio_2_neves2024shedding, radio_3_liao2023deep} and ensures consistency and repeatability in the analysis of medical images. This automation is crucial for handling the large data generated by MRI scans, allowing for quicker and more accurate diagnoses \cite{cheng2021deep}.

This paper focuses on the segmentation of lumbar spine structures, specifically targeting four classes: background, vertebrae, spinal canal, and IVDs. These classes are critical for a comprehensive spine analysis and any pathological changes. By automating the segmentation process, radiologists can allocate more time to critical diagnostic tasks \cite{radio_1_wataya2023radiologists}, reduce human error, and increase throughput in clinical settings \cite{najjar2023redefining}. Automated segmentation supports radiologists by providing quantitative measures that can aid in clinical decision-making \cite{patel2023improving,radio_2_neves2024shedding}.

Despite significant advancements in deep learning, challenges such as class imbalance and the complexity of spine anatomy remain significant hurdles. Addressing these challenges effectively is essential for developing robust and reliable segmentation models \cite{Class_imbalance_better_model_tholke2023class, spi_2petrakis2024lunar}. Related works in lumbar spine segmentation have utilized machine learning and deep learning approaches. However, many of these methods struggle with class imbalance and do not adequately handle the complex anatomy of the lumbar spine.

Using the SPIDER dataset \cite{datasetvan2023spider}, a well-established and publicly available dataset for lumbar spine segmentation in MR images, provides a realistic representation of clinical conditions with MRI scans from patients with low back pain. Initial data preprocessing revealed several issues, including missing vertebrae in 2D slices and an incorrect classification scheme, which initially resulted in 16 classes instead of the intended four.

To overcome these challenges, a comprehensive data preprocessing pipeline was implemented. First, 2D PNG images were extracted from the 3D MRI MetaImage Medical Format (mha) files. The missing vertebrae issue was identified and rectified during this process, ensuring that each 2D image accurately represented the four target classes. Rigorous data filtration techniques were applied to exclude images with fewer than four classes and those exhibiting an excessive class imbalance ratio (greater than 55\%). This step was critical to ensure the quality and reliability of the training data.

A modified U-Net model was employed for the segmentation task, known for its efficacy in medical image segmentation \cite{UNET_2_siddique2021u}. Modifications included the integration of an upsample block with leaky Rectified Linear Units (ReLU) (alpha=0.1) \cite{leaky_relu_1_mastromichalakis2020alrelu,leaky_relu_2_dubey2019comparative} to mitigate the dying ReLU \cite{dying_relu_1_douglas2018relu} problem. In this common issue, neurons become inactive and cease learning. The Glorot uniform initializer was also utilized for kernel weights to stabilize the training process \cite{glorot_1_chumachenko2022feedforward, Glorot_2_aguirre2019improving}. A custom combined loss function was devised to address the class imbalance issue, which is prevalent in medical image segmentation \cite{classimbalance_3_cui2024unified}, incorporating both focal loss and dice loss. This approach not only penalizes the model for misclassifying minority classes but also ensures smooth gradient flow, enhancing overall performance\cite{classimbalance_1_yeung2022unified,Class_imbalance_better_model_tholke2023class}.

Evaluation of the model's performance was conducted using a suite of metrics including Mean Intersection over Union (Mean IoU), Dice coefficient, Average Surface Distance (ASD), Normalized Surface Distance (NSD), precision, recall, and F1 score for each class. These metrics provided a comprehensive analysis of the model's segmentation capabilities, ensuring that each class was accurately identified and delineated \cite{Metric_various_wang2024revisiting,metric_dice_liu2024we,metric_3_opitz2024closer}.

The results demonstrated significant improvements across all metrics compared to existing methods applied to the same dataset. This highlights the efficacy of the data preprocessing techniques, the robustness of the modified U-Net architecture, and the effectiveness of the custom loss function in handling class imbalance.

This study introduces a novel approach to lumbar spine segmentation, effectively addressing key challenges in data preprocessing and model training. Utilizing a carefully curated dataset and a modified deep learning model, the approach offers a robust solution that advances the state-of-the-art in medical image segmentation. The contributions presented have significant implications for clinical applications, with the potential to enhance diagnostic accuracy and treatment planning for patients experiencing low back pain. 

The contributions can be summarized as follows:
\begin{itemize}
    \item Implemented a robust preprocessing pipeline to extract 2D PNG images from 3D MRI mha files, ensuring accurate representation of four classes: background, vertebrae, spinal canal, and IVDs.
    \item Developed a advanced data filtering method in data preprocesing and a custom combined loss function (focal loss + dice loss) to effectively handle class imbalance, improving model performance on minority classes.
    \item Modified U-Net model incorporating an upsample block with leaky ReLU (alpha=0.1) and Glorot uniform initializer to prevent the dying ReLU problem and stabilize training.
    \item Applied strict data filtration criteria, removing images with fewer than four classes and those with a class imbalance ratio greater than 55\%, ensuring high-quality training data.
    \item Used multiple metrics (Mean IoU, Dice coefficient, ASD, NSD, precision, recall, F1 score) to thoroughly assess model performance for each class, demonstrating significant improvements over existing methods.

    \item Achieved superior results across all evaluation metrics compared to existing studies using the same dataset, highlighting the effectiveness of the preprocessing techniques and the modified U-Net model.
\end{itemize}


\section{Related Works}
The segmentation process of the Lumbar Spine involves delineating anatomical structures such as vertebrae, spinal canal, and IVDs in MRI scans. Traditional methods for this task, while pioneering, often struggle with the intricate anatomy and variability of the lumbar spine.

Various deep learning methods have been employed to segment medical images, demonstrating remarkable performance. Among these, the U-Net \cite{unetronneberger2015u} model has been particularly influential, setting a high standard for segmentation tasks. Building on the success of the original U-Net \cite{unetronneberger2015u}, several U-Net-based models, such as Attention U-Net \cite{attention_unet_oktay2018attention}, U-Net++ \cite{unet++zhou2018unet++}, and others, have been developed. These advanced architectures have consistently shown superior performance in the semantic segmentation of medical images, addressing specific challenges and further enhancing segmentation accuracy and efficiency \cite{unetronneberger2015u,unet++zhou2018unet++,attention_unet_oktay2018attention}. Several studies have applied deep learning to lumbar spine segmentation with varying degrees of success. 

Class weights of mask images for the lumbar spine are typically imbalanced, with the background class having a significantly higher weight compared to other classes. Additionally, the class weight of IVDs is much lower than that of vertebrae and the spinal canal. Addressing these imbalances is crucial during data preprocessing and in designing an appropriate loss function. \cite{11van2024lumbar} focused on the nn-UNET model and evaluated it using various metrics such as Dice score, ASD, and SD, achieving class-wise Dice coefficients in the range of 0.84 to 0.93. 
\cite{12wang2022automatic} processed the images to enhance the contrast and sharpness of different tissues and labeled the mask images into three categories, including five vertebral bodies. Their study introduced an improved Attention U-Net, featuring two residual modules replacing the original convolutional blocks. An attention module based on multilevel feature map fusion was employed, and a hybrid loss function was utilized to address class imbalance. The improved Attention U-Net \cite{12wang2022automatic} achieved results with an accuracy of 95.50\%, a recall of 0.9453, and a Dice similarity coefficient of 0.9501. To analyze lumbar spine segmentation more accurately, it is crucial to implement a variety of metrics to assess class-wise segmentation accuracy. Key metrics include Intersection over Union (IoU), Dice coefficient, ASD, NSD, precision, recall, F1-Score and others. In their study, \cite{13silvoster2020efficient} utilized a comprehensive set of these metrics to evaluate segmentation accuracy, achieving impressive results with an accuracy of $99\%$, a recall of 0.90, a precision of 0.89, and a Dice similarity coefficient of 0.924. But, In this study \cite{14peng2021convenient}, Based on the chain structure of the spine, an interactive dual-output vertebrae instance segmentation network was designed to segment the specific vertebrae in CT images, a class-wise evaluation like \cite{11van2024lumbar} was implemented to provide a detailed analysis of segmentation performance. For the intervertebral foramen (IVF) region, the results achieved a Dice coefficient of 0.961 and an ASD of 0.29. For the vertebrae, the results were even more impressive, with a Dice coefficient of 0.968 and an ASD of 0.25. In this study \cite{15van2022segmentation}, an extension of an iterative vertebra segmentation method was utilized, which relies on a 3D fully convolutional neural network to segment the vertebrae one by one. This approach achieved a mean Dice score of $93\%$ $\pm$ 2\% for vertebra segmentation and $86\%$ $\pm 7\%$ for IVDs segmentation. These results demonstrate the efficacy of their method in accurately segmenting vertebrae and IVDs, highlighting its potential for detailed anatomical analysis.  

All the mentioned studies have demonstrated commendable performance in segmenting the lumbar spine. However, none have adequately addressed the issue of class imbalance during the data preprocessing stage. Notably, only \cite{12wang2022automatic} attempted to mitigate class imbalance by incorporating a hybrid loss function. For a comprehensive comparative analysis of lumbar spine segmentation, employing a diverse set of metrics is imperative. While \cite{13silvoster2020efficient} utilized multiple metrics to evaluate segmentation accuracy, they did not conduct a detailed class-wise evaluation. The proposed model addresses the class imbalance issue during the data preprocessing stage and model training by incorporating a combined loss function. This approach ensures superior performance compared to existing studies, providing a more robust and accurate segmentation of lumbar spine structures.

\section{Experiments}
\subsection{Dataset}
The dataset utilized in this study was obtained from the publicly available SPIDER dataset, titled "SPIDER - Lumbar Spine Segmentation in MR Images: A Dataset and a Public Benchmark" \cite{datasetvan2023spider}. This dataset represents a comprehensive multi-center collection of lumbar spine MRI scans, complete with reference segmentations of vertebrae, IVDs, and the spinal canal.

The SPIDER dataset includes many 3D MRI scans, comprising 447 sagittal T1 and T2 MRI series derived from 218 studies of 218 patients, all of whom have a documented history of low back pain. The data was collected across four hospitals, ensuring a diverse representation of imaging practices and patient demographics. This diversity is crucial for developing robust and generalizable segmentation algorithms that aim to promote advancements and collaborative efforts in spine segmentation, ultimately enhancing the diagnostic utility of lumbar spine MRI.
The structured partitioning into training and validation sets and the detailed demographic and imaging metadata provide an excellent foundation for advancing spine segmentation.


\subsection{Data Pre-Processing}
\label{Data Preprocessing}
\subsubsection{Data Extraction}

\begin{figure}[htb!]
  \begin{subfigure}{0.49\columnwidth}
  \includegraphics[width=\textwidth]{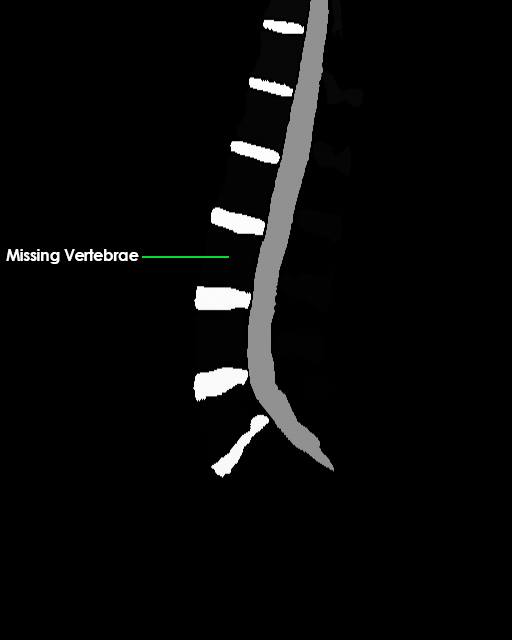}
  \caption{Initial extraction}
  \label{subfig:Missing_Vertebra}
  \end{subfigure}
  \hfill
  \begin{subfigure}{0.49\columnwidth}
  \includegraphics[width=\textwidth]{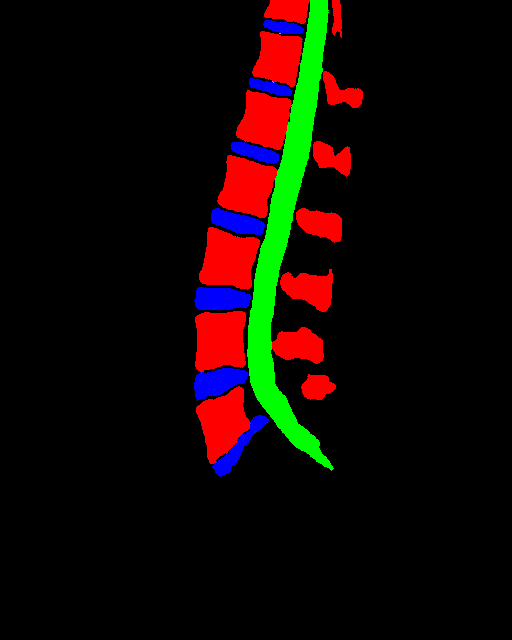}
  \caption{After restoration} 
  \label{subfig:Vertebra_restored}
  \end{subfigure} 
  \caption{Vertebrae Restoration by Custom Data Preprocessing Method}
  \label{fig:Vertebra_missing and restored}
\end{figure}

The default dataset comprised 3D MRI images in mha and corresponding segmentation masks. These images were categorized into three types: T1, T2, and T2\_SPACE. Initial inspection using 3D Slicer \cite{3D_slicer_fedorov20123d} revealed the absence of the vertebrae class from all masks. This issue was resolved by adjusting the color lookup table in 3D Slicer, which made the vertebrae visible. For the deep learning model, the 3D images (Sagittal) were converted into 2D PNG format with a resolution of 512 by 640 pixels, including the MRI images and their corresponding masks.

When extracting 2D PNG images from the 3D mha files, four primary issues needed to be addressed: 
\begin{enumerate}
    \item Vertebrae class is missing
    \item There should be in total 4 classes (Background, Vertebrae, Spinal canal, and IVDs), but 16 classes are present
    \item Some of the slices of the images do not represent the sagittal slices
    \item Some of the images are rotated or flipped
\end{enumerate}
Issue no. 3 was resolved by manually identifying the images that represented the sagittal axis. The axial axis was extracted to obtain the sagittal slices for those specific 3D images in such cases. Issue no. 4 was addressed by rotating or flipping during the extraction process. In issues 1 and 2, the masks again missed the vertebrae. The extracted 2D masks contained 16 classes instead of the expected 4. This discrepancy necessitated further processing steps to correctly represent the vertebrae and reduce the number of courses to the desired four. 

\begin{figure}[htb!]
  \begin{subfigure}{0.32\columnwidth}
  \includegraphics[width=\textwidth]{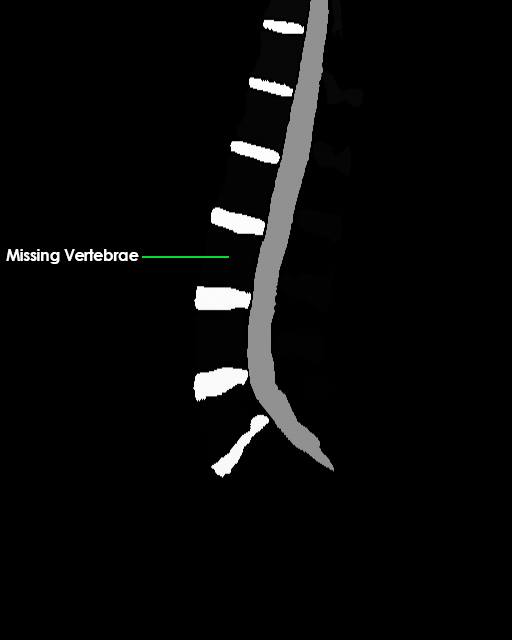}
  \caption{}
  \label{subfig:Data_Extraction_A}
  \end{subfigure}
  \hfill
  \begin{subfigure}{0.32\columnwidth}
  \includegraphics[width=\textwidth]{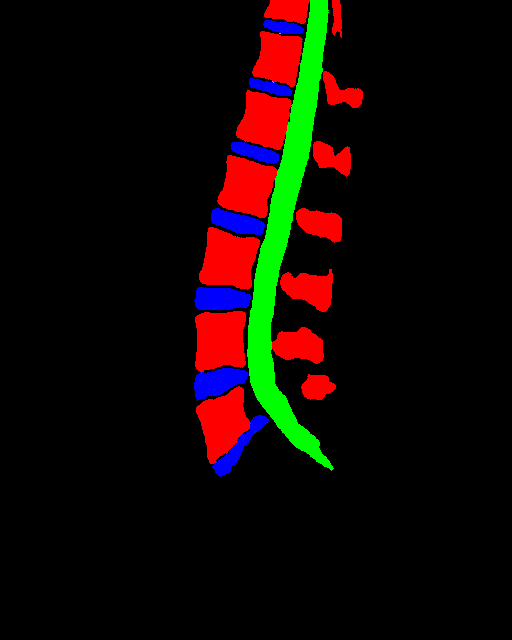}
  \caption{} 
  \label{subfig:Data_Extraction_B}
  \end{subfigure} 
  \hfill
  \begin{subfigure}{0.32\columnwidth}
  \includegraphics[width=\textwidth]{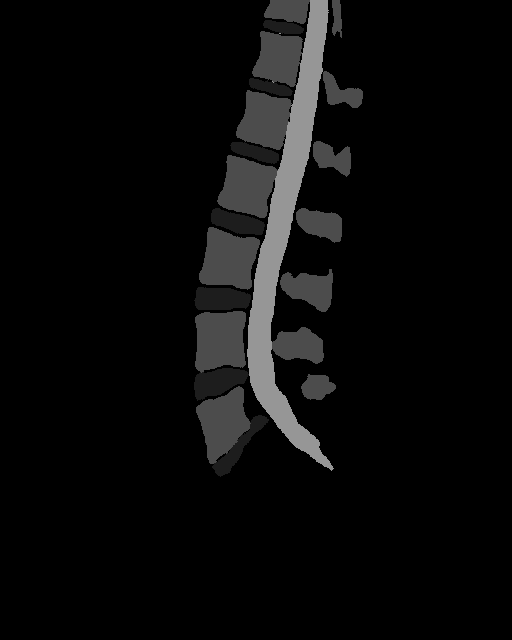}
  \caption{} 
  \label{subfig:Data_Extraction_C}
  \end{subfigure}
  \caption{Data Extraction Process}
  \label{fig:Data_Extraction}
\end{figure}

In Figure \ref{subfig:Missing_Vertebra}, The Vertebra is not visible and the total classes of the image are 16. To overcome these issues, a multi-step preprocessing method was implemented. This method processes the initially extracted images, applying various color transformations and corrections to accurately represent the anatomical structures. It processes initially extracted images (missing vertebrae and 16 classes) from a specified input folder, applying multiple color transformations before saving them to an output folder. Initially, it replaces certain color ranges, such as very dark shades with red, mid-range shades with green, and light shades with blue. It then adjusts each pixel's color to match its neighbors if they share the same color, promoting uniformity. Next, it removes outlines by ensuring pixels that differ significantly from their neighbors adopt the most common surrounding color. The code also specifically addresses border pixels, ensuring their colors align with adjacent pixels.
Additionally, it replaces isolated pixels that stand out with the most common neighboring color. If the image contains green or blue but lacks red, these colors are replaced with red to maintain consistency. The processed images are then saved to the output folder, ensuring each step enhances the visual coherence of the images.

In Figure \ref{subfig:Vertebra_restored}, the missing vertebra issue has been solved. Moreover, it now has exactly four classes, including the background class. Here, the red color represents the restored Vertebrae. Detailed processing steps are following.


Initially, specific color ranges are identified in the extracted images, categorizing them into very dark shades, mid-range shades, and light shades. Dark shades are replaced with red, mid-range shades are replaced with green, and light shades are replaced with blue. This color replacement technique highlights different anatomical structures, making them visually distinct and easier to analyze. Each pixel's color is then adjusted to match its neighboring pixels if they share the same color. This step enhances uniformity across the image and ensures smooth transitions between adjacent regions, reducing noise and enhancing the overall coherence of the image. Outliers, or pixels that differ significantly from their surrounding neighbors are identified next. These outliers are replaced with the most common surrounding color, effectively removing outlines and sharpening the boundaries within the segmentation masks. This process contributes to a cleaner and more precise representation of the structures. Special attention is given to border pixels, ensuring their colors align with those of adjacent pixels. This adjustment prevents discontinuities at the borders, maintaining the integrity of the segmentation masks and ensuring that boundaries are smooth and consistent.
Additionally, isolated pixels that stand out from their surroundings are identified and replaced with the most common neighboring color. This step further enhances the uniformity and coherence of the segmentation masks, eliminating any residual noise or artifacts. In cases where an image contains green or blue but lacks red, these colors are replaced with red to maintain consistency across all images. This ensures that all images in the dataset have a uniform color scheme, which is critical for accurate segmentation. Finally, the processed images are saved to an output folder, preserving the enhancements made during preprocessing. Each step in this pipeline is carefully designed to improve the visual coherence of the images, facilitating more accurate and reliable segmentation, which is essential for downstream medical analysis.

Implementing this enhanced preprocessing method effectively addresses the initial issues encountered during data extraction, ensuring that the segmentation masks accurately represent the vertebrae, spinal canal, and IVDs in lumbar spine MRI images.



\begin{algorithm}[htb!]
\caption{Adaptive Pixel Transformation Algorithm (APTA)}\label{alg:three}

\begin{algorithmic}[1]

\Require Input image $I$
\Ensure Output image $O$

\textbf{\textit{for}} each pixel \( p_{ij} \) at position \( (i, j) \) in \( I \) \textbf{\textit{do}}
\\Apply the following operations to each pixel:
\begin{enumerate}
    \item \textbf{Replace pixel colors falling within specified ranges} 
    \\with red if not already present:
    \begin{enumerate}
        \item Let \( T : \mathbb{Z}^3 \rightarrow \mathbb{Z}^3 \) be the color transformation function defined as \( T(c) = (255, 0, 0) \) for \( c \) in the specified color range.
        \item Apply \( T \) to \( p_{ij} \) if \( p_{ij} \) falls within the specified \\ color range.
    \end{enumerate}
    \item \textbf{Replace \( p_{ij} \) with the color of the first matching adjacent pixel \( p_{xy} \) if \( p_{xy} \) matches \( p_{ij} \)}:
    \begin{enumerate}
        \item Let \( f : \mathbb{Z}^2 \rightarrow \mathbb{Z}^3 \) represent the pixel colors of \( I \).
        \item Define a pixel-wise transformation \( T : \mathbb{Z}^3 \rightarrow \mathbb{Z}^3 \) such that \( T(p_{ij}) = p_{xy} \) if \( p_{xy} \) is the color of the \\first matching adjacent pixel to \( p_{ij} \).
    \end{enumerate}
    \item \textbf{Remove the outline or boundary of objects by replacing \( p_{ij} \) with the most common neighboring color}:
    \begin{enumerate}
        \item Let \( C(p_{ij}) \) denote the multiset of colors of neighboring pixels to \( p_{ij} \).
        \item Replace \( p_{ij} \) with the most common color in \\ \( C(p_{ij}) \).
    \end{enumerate}
    \item \textbf{Replace \( p_{ij} \) with the most common neighboring \\color if both adjacent pixels have different colors}:
    \begin{enumerate}
        \item Let \( f : \mathbb{Z}^2 \rightarrow \mathbb{Z}^3 \) represent the pixel colors of \( I \).
        \item Define a conditional transformation \( T : \mathbb{Z}^3 \rightarrow \mathbb{Z}^3 \) such that \( T(p_{ij}) = \text{MostCommonNeighbor}(p_{ij}) \) if both adjacent pixels to \( p_{ij} \) have different colors.
    \end{enumerate}
    \item \textbf{Replace isolated pixels (singletons) with the color of their most common neighboring pixel}:
    \begin{enumerate}
        \item Define a function \( \text{IsSingleton}(p_{ij}) \) that determines \\ if \( p_{ij} \) is a singleton based on its neighboring \\ pixels.
        \item If \( p_{ij} \) is a singleton, replace it with the color of its most common neighboring pixel.
    \end{enumerate}
    \item \textbf{Replace green or blue pixels with red if red color is \\not already present}:
    \begin{enumerate}
        \item Let \( f : \mathbb{Z}^2 \rightarrow \mathbb{Z}^3 \) represent the pixel colors of \( I \).
        \item Define a conditional transformation \( T : \mathbb{Z}^3 \rightarrow \mathbb{Z}^3 \) such that \( T(p_{ij}) = (255, 0, 0) \) if the specified condition is met.
    \end{enumerate}
\end{enumerate}
Save the modified image $O$\\
\textbf{end \textit{for}}
\end{algorithmic}
\end{algorithm}

In Figure \ref{subfig:Data_Extraction_A}, the 3D sagittal masks were missing vertebrae and included 16 confusing classes, making segmentation challenging for the deep learning model. To address this, a detailed preprocessing method APTA was applied, involving color transformations that restored the vertebrae class and refined the segmentation to 4 classes, as shown in Figure \ref{subfig:Data_Extraction_B}. Finally, in Figure \ref{subfig:Data_Extraction_C}, the RGB images were converted to grayscale, preserving class information while simplifying the data for training and effectively preparing the dataset for filtration.


\subsubsection{Data Filtration}

The data filtration stage focused on refining the extracted images to ensure the dataset's suitability and balance for training the deep learning model. The filtration process involved two main steps: eliminating unnecessary images and addressing class imbalance within the dataset.

\subsection*{Step 1: Elimination of Redundant Images}

 Initially, in the dataset, masks images that contained less than 4 classes, precisely the black background, were deemed irrelevant for the training process. So, the corresponding original images with the mask images (less than four classes) were filtered out. These images do not contribute useful information for the segmentation task and were thus excluded from the dataset. This step ensures that all remaining images contain significant anatomical structures necessary for the training and validation of the deep learning model.

 \begin{figure}[htb!]
  \begin{subfigure}{0.17\columnwidth}
  \includegraphics[width=\textwidth]{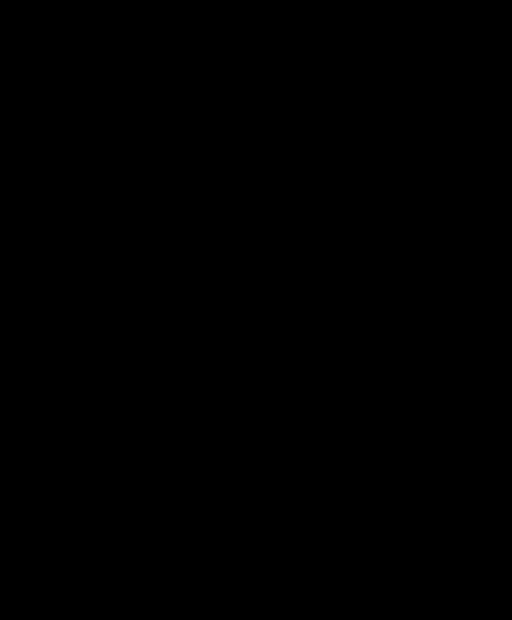}
  \caption{}
  \label{subfig:Elimination_A}
  \end{subfigure}
  \hfill
  \begin{subfigure}{0.17\columnwidth}
  \includegraphics[width=\textwidth]{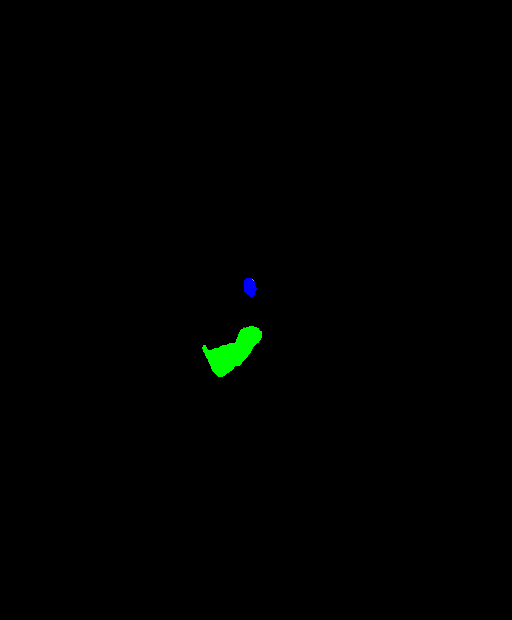}
  \caption{} 
  \label{subfig:Elimination_B}
  \end{subfigure} 
  \hfill
  \begin{subfigure}{0.17\columnwidth}
  \includegraphics[width=\textwidth]{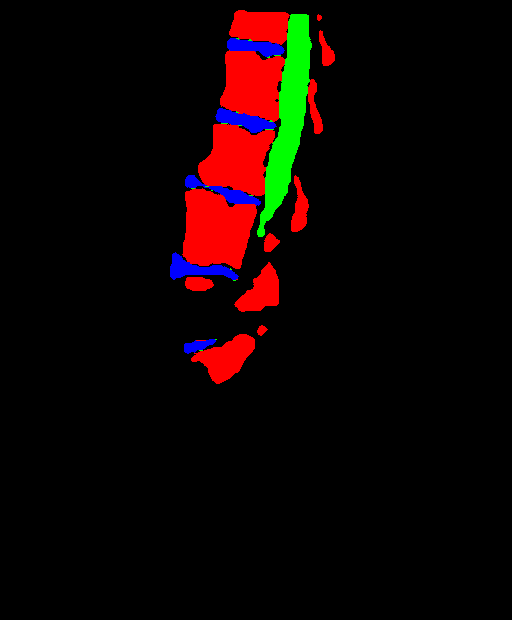}
  \caption{} 
  \label{subfig:Elimination_C}
  \end{subfigure}
\hfill
  \begin{subfigure}{0.17\columnwidth}
  \includegraphics[width=\textwidth]{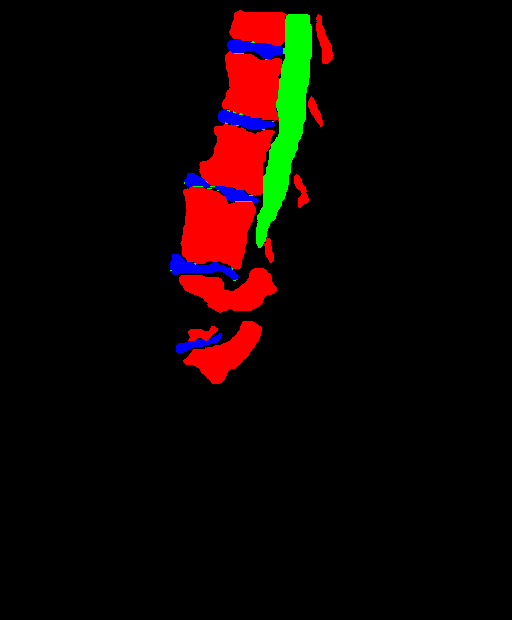}
  \caption{} 
  \label{subfig:Elimination_D}
  \end{subfigure} 
\hfill
  \begin{subfigure}{0.17\columnwidth}
  \includegraphics[width=\textwidth]{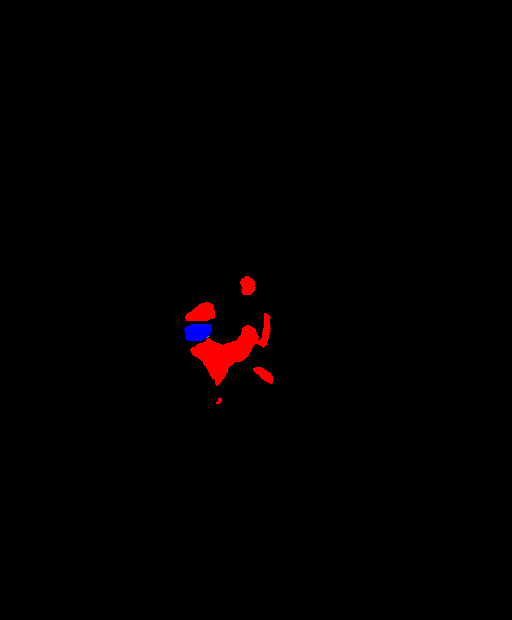}
  \caption{} 
  \label{subfig:Elimination_E}
  \end{subfigure}
  \caption{Elimination of Redundant Images}
  \label{fig:fig:Data_Filtration_Step_1}
\end{figure}

 In Figure \ref{subfig:Elimination_A},  contains only one class and it is the background class, rendering it redundant for the training process. Consequently, this mask image is excluded from the dataset. Similarly, images in Figure \ref{subfig:Elimination_B} \& \ref{subfig:Elimination_E} contain fewer than four classes, which also makes them unsuitable for training and thus they are eliminated. On the other hand, \ref{subfig:Elimination_C} and \ref{subfig:Elimination_D} contain the appropriate number of classes and are therefore deemed suitable for training.

  \begin{figure}[htb!]
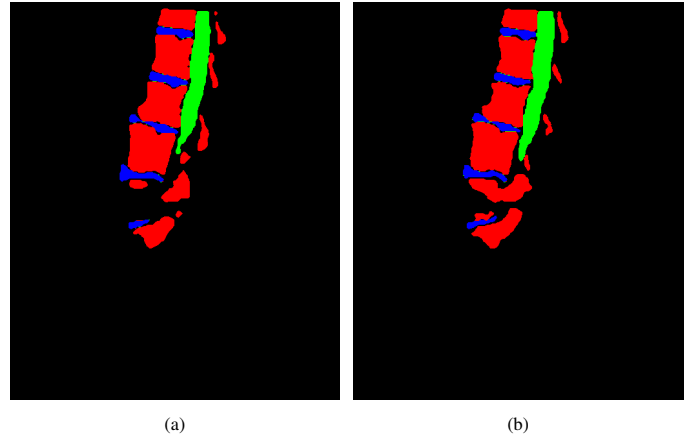

  \begin{subfigure}{0.49\columnwidth}
  \includegraphics[width=\textwidth]{After_Elimination_of_Redundant_Images_C.png}
  \caption{}
  \label{subfig:After_Elimination_of_redundant_Images_C}
  \end{subfigure}
  \hfill
  \begin{subfigure}{0.49\columnwidth}
  \includegraphics[width=\textwidth]{After_Elimination_of_Redundant_Images_D.png}
  \caption{} 
  \label{subfig:After_Elimination_of_redundant_Images_D}
  \end{subfigure} 
  \caption{After Elimination of Redundant Images}
    \label{fig:After_Elimination_of_redundant_Images}
\end{figure}

Finally, the Figure \ref{subfig:After_Elimination_of_redundant_Images_C} and \ref{subfig:After_Elimination_of_redundant_Images_D} are selected by filtering out Figure \ref{subfig:Elimination_A}, Figure \ref{subfig:Elimination_B} and Figure \ref{subfig:Elimination_E}.

\subsection*{Step 2: Addressing Class Imbalance}

 The dataset comprised three types of images: T1, T2, and T2\_SPACE. To address the class imbalance, the class weights of the dataset were calculated. The class weight for each image type was determined as follows:
\[ \text{Class Weight} = \frac{\text{Number of Pixels of a Specific Class}}{\text{Total Number of Pixels in the Image}} \]

 Once the class weights were determined, the class imbalance ratio was calculated by dividing the highest class weight by the lowest class weight across the dataset.
 
The class imbalance ratio was recalculated after filtering by dividing the highest class weight by the lowest class weight:
\[ \text{Class Imbalance Ratio} = \frac{\text{Highest Class Weight}}{\text{Lowest Class Weight}} \]

 Initially, the class imbalance ratios for T1, T2 and T2\_SPACE are respectively 57\%, 56\% and 40\%.

 Images were filtered based on their class weight ratios to mitigate this imbalance. Specifically, images with a class weight ratio above 55\% were considered over-represented and were subsequently removed. This threshold was empirically determined to reduce the over-representation of particular image types without excessively reducing the dataset size.

\begin{figure}[htb!]
  \begin{subfigure}{0.32\columnwidth}
  \includegraphics[width=\textwidth]{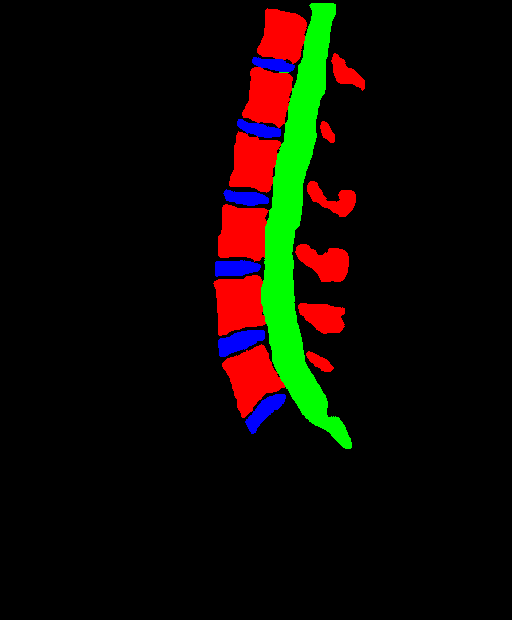}
  \caption{}
  \label{subfig:Handling_Class_Imbalance_A}
  \end{subfigure}
  \hfill
  \begin{subfigure}{0.32\columnwidth}
  \includegraphics[width=\textwidth]{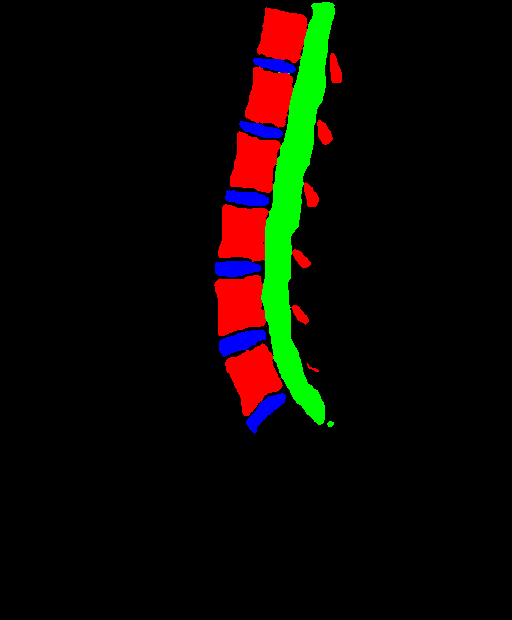}
  \caption{} 
  \label{subfig:Handling_Class_Imbalance_B}
  \end{subfigure} 
  \hfill
  \begin{subfigure}{0.32\columnwidth}
  \includegraphics[width=\textwidth]{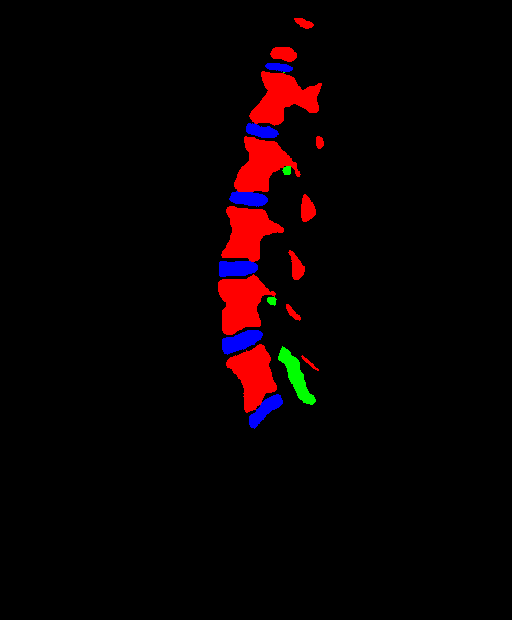}
  \caption{} 
  \label{subfig:Handling_Class_Imbalance_C}
  \end{subfigure}
  \caption{Handling Class Imbalance}
  \label{fig:Handling Class Imbalance}
\end{figure}

 In Figure \ref{fig:Handling Class Imbalance}, Figure \ref{subfig:Handling_Class_Imbalance_C} significantly contributes to class imbalance. Upon calculating the class imbalance ratio for this image, it was found to exceed 55\%. To address this imbalance and improve the dataset's uniformity, Figure \ref{subfig:Handling_Class_Imbalance_C} should be excluded from the training set.

  \begin{figure}[htb!]
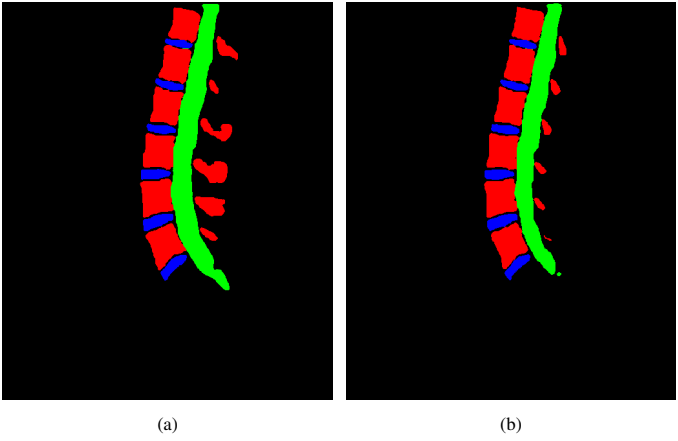

  \begin{subfigure}{0.49\columnwidth}
  \includegraphics[width=\textwidth]{After_Handling_Class_Imbalance_A_237_t2_slice8.png}
  \caption{}
  \label{subfig:After_Handling_A}
  \end{subfigure}
  \hfill
  \begin{subfigure}{0.49\columnwidth}
  \includegraphics[width=\textwidth]{After_Handling_Class_Imbalance_b_237_t2_slice8.png}
  \caption{} 
  \label{subfig:After_Handling_B}
  \end{subfigure} 
  \caption{After Handling Class Imbalance}
    \label{fig:After Handling Class Imbalance}
\end{figure}

 After applying this filtration step, the class distribution was adjusted to 55\% for T1, 55\% for T2, and 37\% for T2\_SPACE images.

In Figure \ref{fig:After Handling Class Imbalance}, Figure \ref{subfig:After_Handling_A} and Figure \ref{subfig:After_Handling_B} are evaluated for their suitability in the training dataset. After thorough analysis and verification, these images meet the required criteria and are deemed suitable for inclusion in the training set. Their inclusion ensures a balanced and representative dataset, which is crucial for the practical training of the model. 
This post-filtration adjustment reduced the class imbalance to 3.51\% for T1, 1.79\% for T2, and 7.5\% for T2\_SPACE images. This reduction in class imbalance is critical for training a robust deep learning model, as it prevents the model from becoming biased towards more frequently occurring classes. By ensuring a more uniform distribution of T1, T2, and T2\_SPACE images, the model can learn features across all image types more effectively, leading to improved segmentation performance.

Summing up the data extraction and filtration processes, the dataset T1, T2, and T3 weighted images each has 1000 2D MRI images with corresponding masks referring in Table \ref{tab:Table_2}.

\begin{table*}[htb!]
\centering
\renewcommand{\arraystretch}{1.4}
\setlength{\tabcolsep}{8.5pt}
\begin{NiceTabular}{lcc}
\CodeBefore
 \rowlistcolors{2}{,tab2}[restart,cols={1-3}]
\Body 
\arrayrulecolor{steelblue}\toprule
        \textbf{Dataset}  & \textbf{No. of Images} & \textbf{Class Imbalance Ratio} \\
\arrayrulecolor{steelblue}\toprule 
     $T1$ & 1000 & 55\%  \\
     $T2$ & 1000 & 55\%  \\
     $T2\_SPACE$ & 1000 & 37\%  \\
     $Total$ & 3000 & ---  \\
\arrayrulecolor{steelblue}\bottomrule
    \end{NiceTabular}
    \caption{Dataset Summary with Class Imbalance Ratio}
\label{tab:Table_2}
\end{table*}



 Overall, the data filtration process improved the quality and balance of the dataset, thereby enhancing the potential accuracy and generalizability of the deep learning model for lumbar spine segmentation.\\


 \subsection{Evaluation Metrics}

 Several key metrics were utilized to evaluate the performance of the modified U-Net model comprehensively on the task of lumbar spine segmentation. These metrics provide a holistic view of the model's accuracy, precision, and ability to handle class imbalance. Each metric and its mathematical formulation are detailed below:

\subsection*{Mean Intersection over Union (Mean IoU)}
The IoU measures the overlap between the predicted segmentation and the ground truth. For each class \(c\), IoU is defined as:
\[
\text{IoU}_c = \frac{|P_c \cap G_c|}{|P_c \cup G_c|}
\]
where \(P_c\) and \(G_c\) are the predicted and ground truth binary masks for class \(c\), respectively. Mean IoU is the average IoU across all classes:
\[
\text{Mean IoU} = \frac{1}{C} \sum_{c=1}^C \text{IoU}_c
\]
where \(C\) is the total number of classes.

\subsection*{Dice Coefficient}
The Dice coefficient, also known as the Sørensen-Dice index, is another measure of overlap between the predicted and ground truth segmentations. For each class \(c\), the Dice coefficient is defined as:
\[
\text{Dice}_c = \frac{2|P_c \cap G_c|}{|P_c| + |G_c|}
\]
The average Dice coefficient is calculated across all classes.

\subsection*{Average Surface Distance (ASD)}
The ASD quantifies the average distance between the surfaces of the predicted segmentation and the ground truth. For a given class \(c\), ASD is computed as:
\[
\text{ASD}_c = \frac{1}{|S_p| + |S_g|} \left( \sum_{x \in S_p} \min_{y \in S_g} d(x, y) + \sum_{y \in S_g} \min_{x \in S_p} d(y, x) \right)
\]
where \(S_p\) and \(S_g\) are the sets of surface points in the predicted and ground truth segmentations, respectively, and \(d(x, y)\) is the Euclidean distance between points \(x\) and \(y\).

\subsection*{Normalized Surface Distance (NSD)}
NSD measures the surface distance normalized by a tolerance value \(\tau\):
\[
\text{NSD}_c = \frac{1}{|S_g|} \sum_{y \in S_g} \mathbb{1}\left[\min_{x \in S_p} d(y, x) < \tau \right]
\]
where \(\mathbb{1}\) is the indicator function that counts the number of ground truth surface points within the tolerance \(\tau\) from any predicted surface point.

\subsection*{Precision}
Precision for each class \(c\) measures the proportion of true positive predictions among all predicted positives:
\[
\text{Precision}_c = \frac{TP_c}{TP_c + FP_c}
\]
where \(TP_c\) and \(FP_c\) are the true positive and false positive counts for class \(c\), respectively.

\subsection*{Recall}
Recall for each class \(c\) measures the proportion of true positive predictions among all actual positives:
\[
\text{Recall}_c = \frac{TP_c}{TP_c + FN_c}
\]
where \(FN_c\) is the false negative count for class \(c\).

\subsection*{F1 Score}
The F1 score is the harmonic mean of precision and recall for each class \(c\):
\[
F1_c = 2 \cdot \frac{\text{Precision}_c \cdot \text{Recall}_c}{\text{Precision}_c + \text{Recall}_c}
\]
The overall F1 score is averaged across all classes.

These metrics collectively provide a detailed evaluation of the model’s performance, ensuring that each class is accurately segmented and that the model's predictions align closely with the ground truth annotations. The modified U-Net model demonstrates significant improvements across all these metrics compared to existing methods, underscoring the approach's effectiveness in lumbar spine segmentation.
\section{Methodology}

 \begin{figure*}[htb!]
    \centering
    \includegraphics[width=1\linewidth]{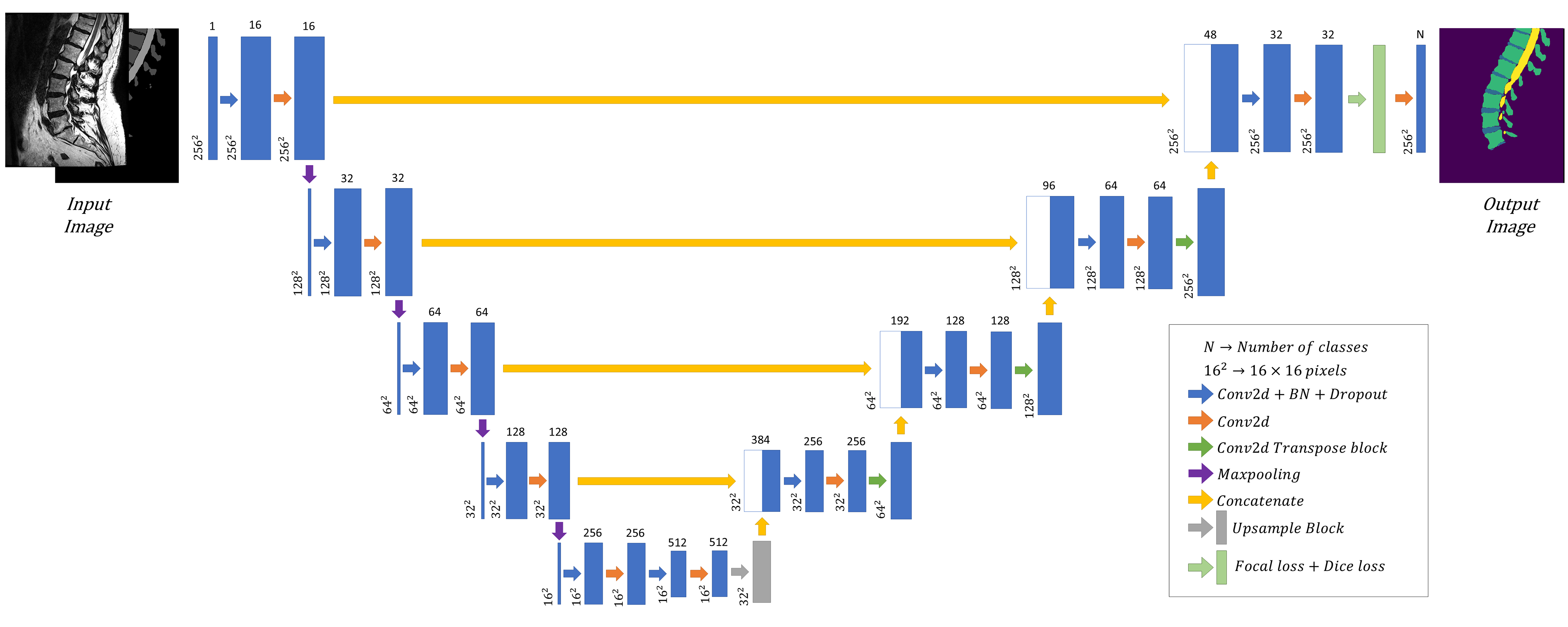}
    \caption{Detailed Architecture of Modified U-Net Model with Custom Upsample Block for Lumbar Spine Segmentation}
    \label{Model_Architecture)}
\end{figure*}

The model architecture in Figure \ref{Model_Architecture)} is based on a modified UNET framework tailored for semantic segmentation tasks in Lumbar Spine segmentation. The model has been meticulously crafted through extensive empirical refinement to optimize its performance for medical image segmentation tasks. It consists of a contraction path followed by an expansive path, and central to its design is an innovative upsample block featuring Leaky ReLU activation.

The model's contractive path begins with convolutional and pooling layers to extract hierarchical features from the input image, with batch normalization and dropout layers included to enhance generalization and prevent overfitting \cite{Batch_Normalization_santurkar2018does, dropout_srivastava2014dropout}. In the expansive path, upsample blocks with \texttt{Conv2DTranspose} layers and Leaky ReLU activation ($\alpha = 0.1$) maintain a small, non-zero gradient even when units are inactive, improving gradient flow and preventing neuron death during training\cite{Leaky_hu2021handling}. An additional 512-channel layer boosts the model's ability to capture intricate details crucial for accurate segmentation. To address the dying ReLU issue, the Glorot uniform initializer is used, promoting stable gradient propagation \cite{golorot_aguirre2019improving}. Addressing the challenge of class imbalance crucial in medical imaging, a bespoke combined loss function combines Focal Loss and Dice Loss, prioritizing hard-to-classify examples while balancing segmentation precision\cite{focal_lin2017focal,dice_zhao2020rethinking}. A combined loss function is adopted, integrating Focal Loss and Dice Loss.

        \[
        L_{\text{focal}}(y_{\text{true}}, y_{\text{pred}}) = -\sum_i \alpha_i (1 - y_{\text{pred}, i})^\gamma y_{\text{true}, i} \log(y_{\text{pred}, i})
        \]
        Focal loss is designed to address class imbalance by focusing more on hard-to-classify examples, reducing the contribution of easily classified examples.

        \[
        L_{\text{dice}}(y_{\text{true}}, y_{\text{pred}}) = 1 - \frac{2 \sum_i y_{\text{true}, i} y_{\text{pred}, i} + \epsilon}{\sum_i y_{\text{true}, i} + \sum_i y_{\text{pred}, i} + \epsilon}
        \]
        Dice loss is a measure of overlap between the predicted segmentation and the ground truth, directly optimizing for the Dice coefficient.
    
    \noindent The Focal Loss component, with a gamma ($\gamma = 4.0$) value of 4.0, emphasizes accurate classification of challenging examples, crucial in medical images where anomalies may be subtle yet significant. Prioritizing performance criteria relevant to medical image analysis, the combined loss function's alpha ($\alpha =.6$) parameter of 0.6 strikes a balance between segmentation precision and minimizing class imbalance. 

        \[
    L_{\text{combined}} = \alpha L_{\text{focal}} + (1 - \alpha) L_{\text{dice}}
    \]

    \noindent By combining focal and dice loss, the model benefits from both approaches.
Alpha ($\alpha$) and Gamma ($\gamma$) Values of the combined loss function were set to  0.6 and 0.4, respectively. The gamma value controls the focus on hard-to-classify regions, and the alpha value controls the balance between focal loss and dice loss functions. This balance ensures that both class imbalance and segmentation accuracy are improved simultaneously\cite{focal_lin2017focal, dice_zhao2020rethinking}.

The modified model can capture more complex features and improve training dynamics. Incorporating focal and dice loss, the combined loss function effectively addresses class imbalance and boundary accuracy. At the same time, early stopping and model checkpointing based on validation Mean IoU ensures robust and generalized performance. These improvements collectively lead to better segmentation results, making the modified model superior for lumbar spine segmentation in MRI scans.

\section{Result \& Analysis}
This section presents the results of the proposed model for the semantic segmentation of lumbar spine MRI scans. The model's performance is evaluated using accuracy, loss, Dice coefficient, Mean IoU, and class-specific metrics.


 \begin{figure}[htb!]
    \centering
    \includegraphics[width=1\linewidth]{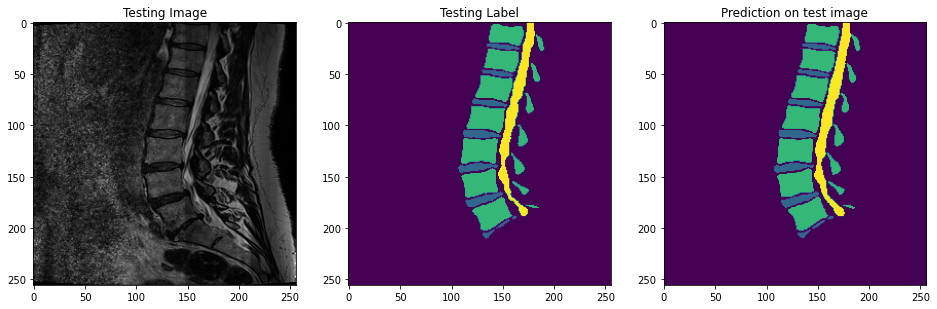}
    \caption{Comparison between Labelled Mask and Predicted Mask}
    \label{Labelled Mask vs Predicted Mask}
\end{figure}

Figure \ref{Labelled Mask vs Predicted Mask} visually compares the predicted segmentation masks and the labeled ground truth masks, referring to the original MRI images. This comparison highlights the model's ability to segment the IVDs, Vertebrae, and Spinal Canal accurately.

 \begin{figure}[htb!]
    \centering
    \includegraphics[width=1\linewidth]{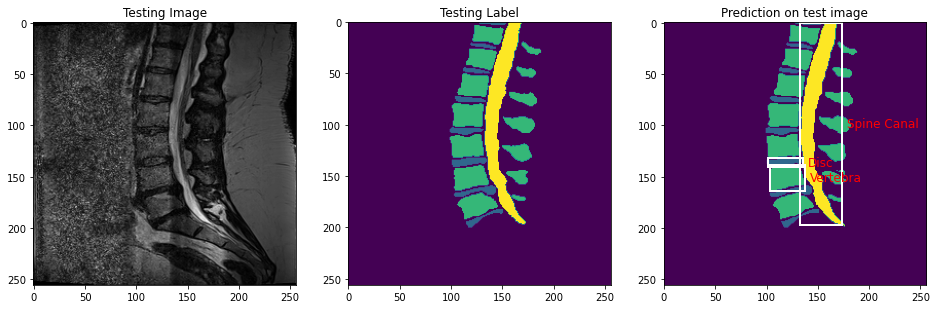}
    \caption{Detection of IVDs, Vertebrae and Spinal Canal in Predicted Mask}
    \label{Class Detection}
\end{figure}

In Figure \ref{Class Detection}, The model effectively distinguished each segment, showcasing its precision in accurately delineating the IVDs, Vertebrae, and Spinal Canal. This contrast underscores the model's proficiency in precisely segmenting these anatomical structures.

\begin{figure}[htb!]
    \centering
    \includegraphics[width=1\linewidth]{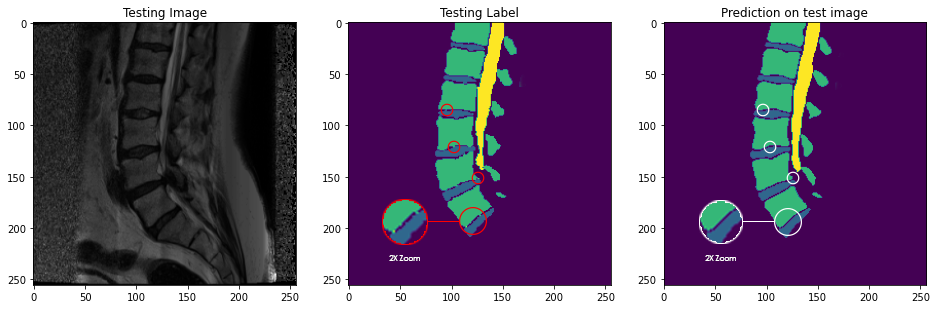}
    \caption{Visualization of Predicted Mask Overlaid on Original Image}
    \label{Detailed_Performance_Analysis}
\end{figure}

In Figure \ref{Detailed_Performance_Analysis}, the red circles on the testing label image highlight areas where the segmentation failed to correctly separate the vertebrae from the IVDs, resulting in noticeable misclassifications. However, in the prediction image marked by white circles, the proposed model demonstrates its effectiveness by successfully resolving these issues. The model accurately separates the vertebrae from the IVDs and corrects previous misclassifications, showcasing its superior performance. The overall accuracy and reliability of the model are clearly evident in these results. 

\begin{figure}[htb!]
    \centering
    \includegraphics[width=1\linewidth]{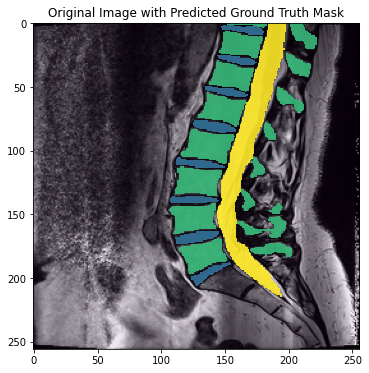}
    \caption{Visualization of Predicted Mask Overlaid on Original Image}
    \label{Visualization of Predicted Mask Overlaid on Original Image}
\end{figure}

Figure \ref{Visualization of Predicted Mask Overlaid on Original Image} demonstrates that the predicted mask aligns seamlessly with the expected positions of each class. The overlap highlights the model's ability to accurately outline the IVDs, Vertebrae, and Spinal Canal. This observation highlights the model's skill in segmenting these anatomical structures with accuracy. 

\begin{figure}[htb!]
    \centering
    \includegraphics[width=1\linewidth]{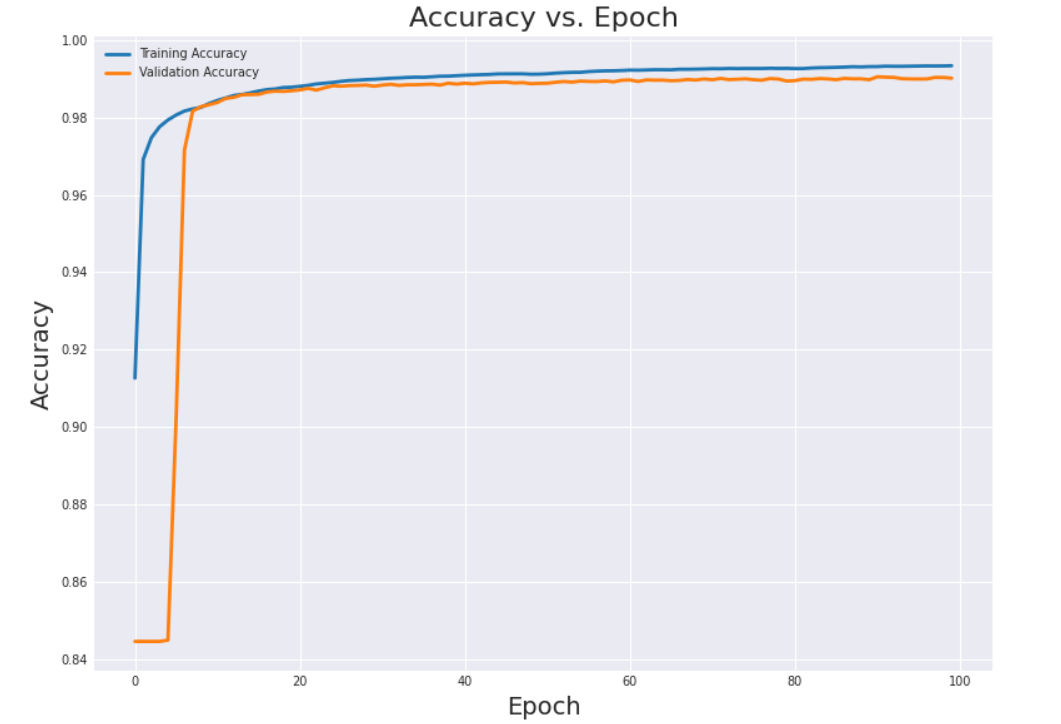}
    \caption{Accuracy vs Epochs}
    \label{fig:Accuracy_vs_Epochs}
\end{figure}



\begin{figure}[htb!]
    \centering
    \includegraphics[width=1\linewidth]{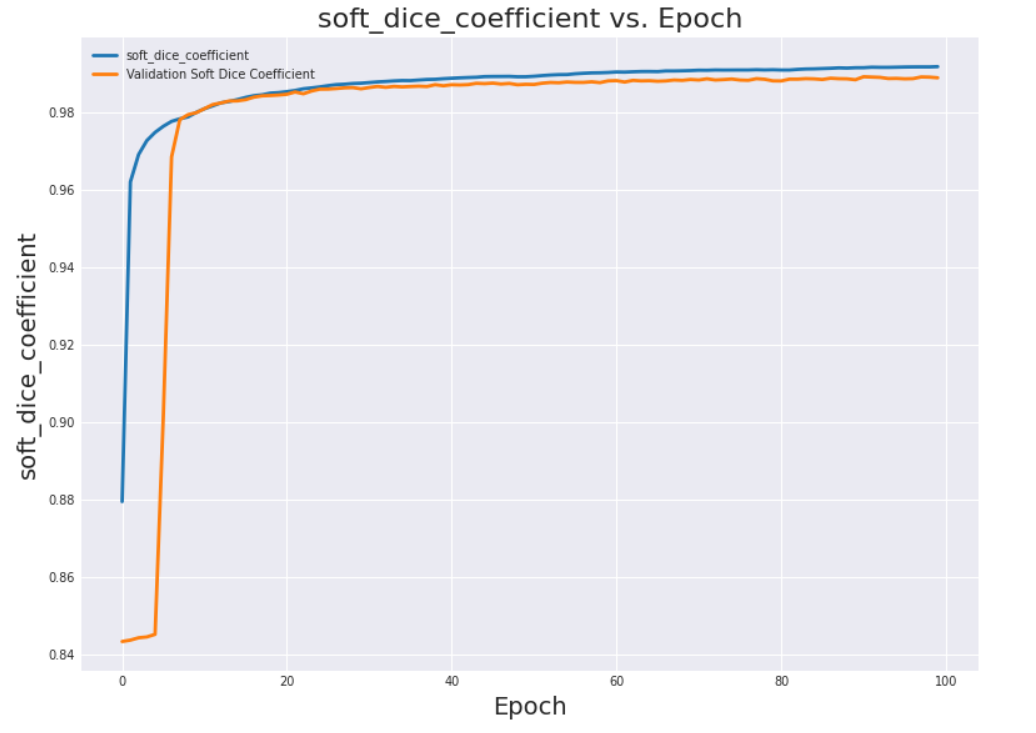}
    \caption{Dice Coefficient vs Epochs}
    \label{fig:Dice Coefficient vs Epochs}
\end{figure}


\begin{figure}[htb!]
    \centering
    \includegraphics[width=1\linewidth]{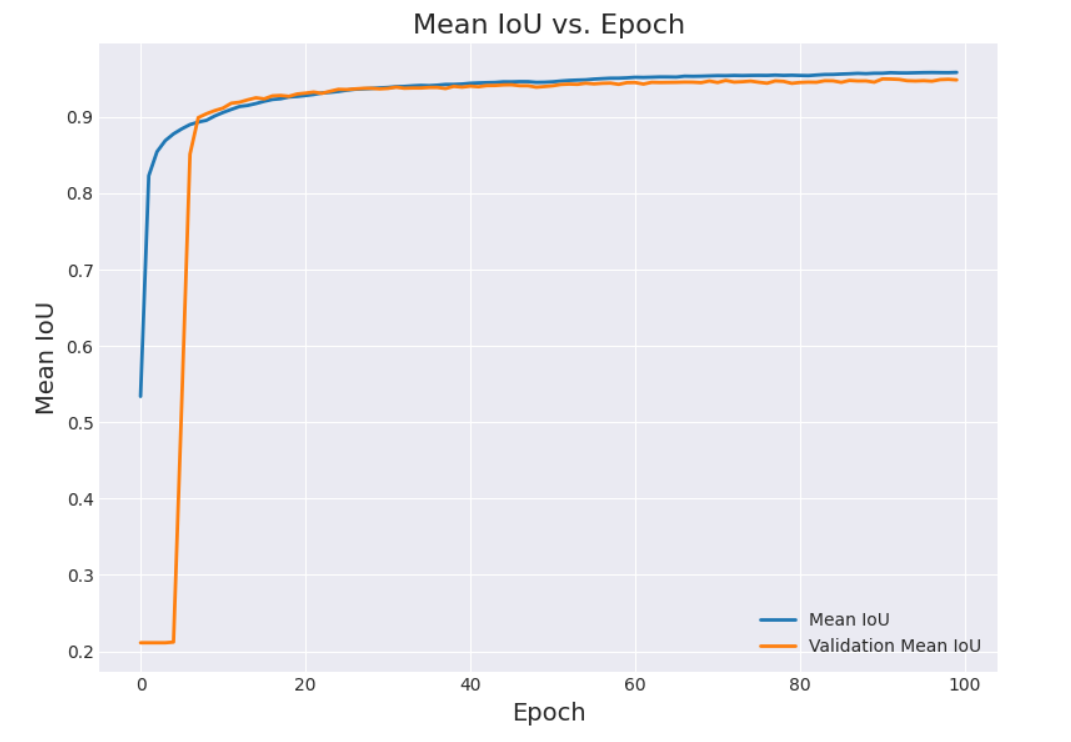}
    \caption{Mean IoU vs Epochs}
    \label{fig:Mean IoU vs Epochs}
\end{figure}

The training progress of the model is illustrated in Figures \ref{fig:Accuracy_vs_Epochs}, \ref{fig:Dice Coefficient vs Epochs},  and \ref{fig:Mean IoU vs Epochs}, which show the Accuracy, Dice coefficient and Mean IoU over 100 epochs. The batch size was set to 8. The training and validation curves almost overlap after 100 epochs, indicating that the model does not suffer from overfitting. This is crucial for ensuring that the model generalizes well to unseen data, which is essential for practical deployment in clinical settings. The absence of overfitting ensures that the model maintains its high performance across different datasets, making it a reliable tool for clinical applications.

\begin{table*}[htb!]
\centering
\renewcommand{\arraystretch}{1.5}
\setlength{\tabcolsep}{8pt}
\begin{NiceTabular}{ccccccccc}
\CodeBefore
   \rowlistcolors{2-13}{tab1,tab2,tab3,tab4}[restart,cols={2-9}]
\Body 
\arrayrulecolor{cite-blue}\toprule
        \textbf{Dataset} & \textbf{Class} & \textbf{IoU $\uparrow$} & \textbf{Dice Coefficient $\uparrow$} & \textbf{ASD $\downarrow$} & \textbf{NSD $\uparrow$} & \textbf{Precision $\uparrow$} & \textbf{Recall $\uparrow$} & \textbf{F1 Score $\uparrow$} \\
\arrayrulecolor{cite-blue}\toprule

\Block{4-1}{$T2\_SPACE$} & Background   & 0.9906 & 0.9950 & 0.0543 & 0.9951 & 0.9951 & 0.9955 & 0.9953 \\
& IVDs         & 0.9476 & 0.9688 & 0.0288 & 0.9994 & 0.9684 & 0.9778 & 0.9731 \\
& Vertebrae    & 0.9461 & 0.9712 & 0.0464 & 0.9944 & 0.9745 & 0.9701 & 0.9723 \\
& Spinal Canal & 0.9501 & 0.9671 & 0.0361 & 0.9963 & 0.9761 & 0.9727 & 0.9744 \\
\hline
\Block{4-1}{$T1$} & Background   & 0.9891 & 0.9893 & 0.0578 & 0.9901 & 0.9899 & 0.9921 & 0.9917 \\
& IVDs         & 0.9378 & 0.9592 & 0.0389 & 0.9954 & 0.9556 & 0.9696 & 0.9663 \\
& Vertebrae    & 0.9367 & 0.9634 & 0.0489 & 0.9912 & 0.9711 & 0.9697 & 0.9678 \\
& Spinal Canal & 0.9445 & 0.9628 & 0.0372 & 0.9893 & 0.9711 & 0.9692 & 0.9725 \\
\hline
\Block{4-1}{$T2$} & Background   & 0.9867 & 0.9877 & 0.0533 & 0.9921 & 0.9856 & 0.9933 & 0.9867 \\
& IVDs         & 0.9317 & 0.9527 & 0.0401 & 0.9934 & 0.9524 & 0.9658 & 0.9650 \\
& Vertebrae    & 0.9331 & 0.9647 & 0.0488 & 0.9919 & 0.9703 & 0.9671 & 0.9666 \\
& Spinal Canal & 0.9439 & 0.9625 & 0.0371 & 0.9856 & 0.9689 & 0.9697 & 0.9709 \\
 \arrayrulecolor{cite-blue}\bottomrule
    \end{NiceTabular}
\caption{Comprehensive Performance Metrics across each Class and Dataset}
    \label{tab:Performance metrics class wise}
\end{table*}
\begin{table*}[htb!]
\centering
\renewcommand{\arraystretch}{1.5}
\setlength{\tabcolsep}{10pt}
\begin{NiceTabular}{cccc}
\CodeBefore
   \rowlistcolors{2-5}{tab1,tab2,tab3,tab4}[restart, cols={2-4}]
   \rowlistcolors{6-11}{tab1,tab2,tab3}[restart, cols={2-4}]
\Body 
\arrayrulecolor{cite-blue}\toprule
        \textbf{Class}     & \textbf{Metric}       & \textbf{Proposed} & \textbf{nn-UNET\cite{11van2024lumbar}}  \\
\arrayrulecolor{cite-blue}\toprule

\Block{4-1}{IVDs} & Mean Dice Coefficient & \textbf{0.9688} & 0.86\\
& IoU & \textbf{0.9476} & --- \\  
& ASD & \textbf{0.0288} & 0.58 \\ 
& NSD & \textbf{0.9994} & ---  \\ 
\hline
\Block{3-1}{Vertebrae}    & Mean Dice Coefficient & \textbf{0.9712} & 0.92 \\ 
& ASD                   & \textbf{0.0464}               & 0.51                  \\ 
& Detection             & \textbf{100\%}                & 99.7\%                \\ 
\hline
\Block{3-1}{Spinal Canal} & Mean Dice Coefficient & \textbf{0.9671}               & 0.92                  \\ 
                                   & ASD                   & \textbf{0.0361}               & 0.46                  \\ 
                                   & Detection             & \textbf{100\%}                & 100\%                 \\
 \arrayrulecolor{cite-blue}\bottomrule
    \end{NiceTabular}
\caption{Performance Comparison of Proposed Model vs. nn-UNET Across Multiple Segmentation Metrics for Lumbar Spine MRI}
    \label{tab:model_comparison}
\end{table*}

A comprehensive comparison was conducted between the proposed model, and another model, nn-UNET\cite{11van2024lumbar}, using the same dataset. Notably, the evaluation of the nn-UNET\cite{11van2024lumbar} model lacked several crucial metrics essential for medical image segmentation tasks, such as IoU, NSD, Precision, Recall, and F1-score. In contrast, the proposed model was assessed using a comprehensive set of metrics necessary for such tasks.

As illustrated in Table\ref{tab:model_comparison}, the scores obtained by the proposed model significantly outperform those of nn-UNET\cite{11van2024lumbar} across all metrics. This clear disparity underscores the superior performance of the proposed model model in the segmentation task. Specifically, the mean Dice coefficient, a widely used metric for assessing segmentation accuracy, was notably higher for the proposed model compared to nn-UNET\cite{11van2024lumbar} in all classes, including IVDs, Vertebrae, and the Spinal Canal. This indicates that the model achieved better overlap and boundary delineation between the predicted and ground truth masks, leading to more accurate segmentation.

Furthermore, the proposed model exhibited lower ASD values for all classes, implying superior boundary detection and closer alignment with the actual anatomical structures. The high NSD values obtained by the proposed model also suggest better overall alignment and spatial correspondence between the segmented regions and the ground truth masks.

While the nn-UNET\cite{11van2024lumbar} model achieved comparable detection rates for some classes, the lack of comprehensive evaluation metrics limits its utility in assessing segmentation quality. In contrast, the detailed evaluation provided by the proposed model, including IoU, NSD, Precision, Recall, and F1-score, offers a more comprehensive understanding of its segmentation performance.

Overall, the results highlight the effectiveness of the proposed model model in accurately segmenting lumbar spine structures from MRI images. Its superior performance, as demonstrated by a comprehensive range of metrics, makes it a promising tool for clinical applications, including disease diagnosis, treatment planning, and medical research.
\section{Conclusion}
 This study presents a sophisticated approach to lumbar spine segmentation using deep learning techniques, with a focus on addressing key challenges such as class imbalance and data preprocessing. By meticulously preprocessing MRI scans of patients with low back pain and rectifying class inconsistencies, the fidelity of the training data is ensured, laying the foundation for an accurate representation of critical classes: vertebrae, spinal canal, and IVDs. The highly modified U-Net model (proposed model), incorporating innovative architectural enhancements and a custom combined loss function, effectively tackles class imbalance and improves segmentation accuracy. Evaluation using a comprehensive suite of metrics demonstrates the superior performance of this approach, outperforming existing methods and advancing current techniques in lumbar spine segmentation. The comparison with another model nn-UNET, highlights the superiority of the proposed model, with significantly higher scores across all evaluation metrics. Specifically, the proposed method achieves better overlap, boundary delineation, and alignment with ground truth masks, making it a promising tool for clinical applications such as disease diagnosis, treatment planning, and medical research.

Overall, the contributions of this study hold significant implications for enhancing diagnostic accuracy and treatment planning for patients with low back pain, showcasing the potential of deep learning techniques in advancing medical image segmentation.
\section*{Acknowledgment}
Istiak Ahmed and Md. Zahirul Islam Nahid would like to thank Dr. Mohammad Monirujjaman Khan and Dr. Mohammad Monir Uddin for their guidance during their CSE499 undergraduate project work at North South University. Some of the results presented in this paper were derived from that project.
\bibliographystyle{elsarticle-num} 
\bibliography{ref}

\end{document}